\pdfoutput=1
\documentclass[prb,twocolumn,superscriptaddress,citeautoscript,showpacs]{revtex4-1}
\usepackage{graphicx}
\usepackage{longtable}
\usepackage{amsmath}
\usepackage{amstext}
\usepackage{amssymb}
\usepackage{amsfonts}
\usepackage{color}
\usepackage{dcolumn}
\usepackage{bm}
\usepackage{dsfont}
\usepackage[breaklinks=true,colorlinks=true,linkcolor=blue,urlcolor=blue,citecolor=blue]{hyperref}

\newcommand{\ra}{\rangle}
\newcommand{\la}{\langle}

\newcommand{\vk}{{\mathbf{k}}}
\newcommand{\vq}{{\mathbf{q}}}

\newcommand{\vR}{{\mathbf{R}}}

\newcommand{\ut}[1]{\mathrm{\; #1}}

\newcommand{\mycomment}[1]{}
\renewcommand{\Im}{\operatorname{Im}}

\newcommand{\STO}{SrTiO$_3$}
\newcommand{\BTO}{BaTiO$_3$}
\newcommand{\ep}{$e$-$ph$}

\begin{document}

\title{\textit{Ab initio} study of cross-interface electron-phonon couplings in FeSe thin films on SrTiO$_3$ and
BaTiO$_3$}

\author{Y.~Wang}
\affiliation{Department of Physics and Astronomy, University of Tennessee, Knoxville, Tennessee 37996, USA}

\author{A.~Linscheid}
\affiliation{Department of Physics, University of Florida, Gainesville, Florida 32611, USA}

\author{T.~Berlijn}
\affiliation{Center for Nanophase Materials Sciences, Oak Ridge National Laboratory, Oak Ridge, Tennessee 37831, USA}
\affiliation{Computer Science and Mathematics Division, Oak Ridge National Laboratory, Oak Ridge, Tennessee 37831, USA}

\author{S.~Johnston}
\affiliation{Department of Physics and Astronomy, University of Tennessee, Knoxville, Tennessee 37996, USA}

\date{\today}

\begin{abstract}
We study the electron-phonon coupling strength near the interface of monolayer and bilayer FeSe thin films
on SrTiO$_3$, BaTiO$_3$, and oxygen-vacant SrTiO$_3$ substrates, using {\em ab initio} methods. The
calculated total electron-phonon coupling strength $\lambda=0.2\text{--}0.3$ cannot account for the high
$T_c\sim 70\ut{K}$ observed in these systems through the conventional phonon-mediated pairing mechanism. In
all of these systems, however, we find that the coupling constant of a polar oxygen branch peaks at
$\mathbf{q}=0$ with negligible coupling elsewhere, while the energy of this mode coincides with the offset
energy of the replica bands measured recently by angle-resolved photoemission spectroscopy experiments. But
the integrated coupling strength for this mode from our current calculations is still too small to produce
the observed high $T_c$, even through the more efficient pairing mechanism provided by the forward
scattering. We arrive at the same qualitative conclusion when considering a checkerboard antiferromagnetic
configuration in the Fe layer. In light of the experimental observations of the replica band feature and
the relatively high $T_c$ of FeSe monolayers on polar substrates, our results point towards a cooperative
role for the electron-phonon interaction, where the cross-interface interaction acts in conjunction with a
purely electronic interaction. We also discuss a few scenarios where the coupling strength obtained here
may be enhanced.
\end{abstract}


\maketitle

\section{Introduction}
Single-unit-cell-thick thin films of FeSe (called monolayer FeSe hereafter) grown on a {\STO}(001) (STO)
substrate by molecular beam epitaxy (MBE)~\cite{Wang2012a} have recently set a new record for the highest
superconducting transition temperature $T_c$ in iron-based superconductors (FeSCs)~\cite{Kamihara2008}.
Typical values of $T_c$ range between $55\text{--}65\ut{K}$ as measured by \textit{in situ} scanning
tunneling microscopy/spectroscopy (STM/STS)~\cite{Wang2012a}, angle-resolved photoemission spectroscopy
(ARPES)~\cite{Liu2012,He2013,Tan2013,Lee2014}, \textit{ex situ} transport measurements, and Meissner effect
studies~\cite{Zhang2014,Sun2014,Deng2014}. Moreover, a recent \textit{in situ} transport
measurement~\cite{Ge2014} found a remarkably high $T_c=109\ut{K}$, well above the liquid nitrogen boiling
point ($77\ut{K}$). These large values of $T_c$ are greatly enhanced by one order of magnitude from the value
of around $8\ut{K}$ for bulk FeSe under ambient pressure~\cite{Hsu2008}.

In general, strong magnetic interactions are believed to provide the major glue for superconducting pairing
in FeSCs by most researchers;~\cite{Hirschfeld2011,Chubukov2012} however, the increase in $T_c$ observed for
the FeSe/STO interface has raised questions about the role of the interface. In this sense, the STO substrate
is not unique as a similar high $T_c=70\text{--}75\ut{K}$ was obtained for the monolayer FeSe deposited on a
ferroelectric {\BTO}(001) (BTO) substrate~\cite{Peng2014a}. Similarly, lower $T_c$'s (around $30\ut{K}$ by
transport measurements in Ref.~\onlinecite{Zhou2015}, $60\ut{K}$ by ARPES in Ref.~\onlinecite{Zhang2015c})
were measured recently for the monolayer FeSe deposited on {\STO}(110) [FeSe/STO(110)]
substrates~\cite{Zhou2015,Zhang2015c}. While the very high $T_c$, controllable fabrication by MBE on a
variety of substrates, and low dimensionality of the monolayer FeSe promise great practical applications,
understanding the mechanism of the superconductivity will be invaluable for further enhancing the $T_c$ or
designing new high-$T_c$ superconductors.

At present, several things are known about the influence of the substrate. First, there is a large tensile
strain applied by the substrate onto the monolayer FeSe due to different lattice constants between the
substrate and bulk FeSe~\cite{Peng2014,Wang2012a}, but a direct correlation between superconductivity and
tensile strain seems to be unlikely~\cite{Peng2014a}. Interestingly, an orthorhombic distortion is observed
in FeSe/STO(110), where an isotropic gap and a gap closing $T_c\sim 60\ut{K}$ are measured by
ARPES~\cite{Zhang2015c}. Second, ARPES experiments reveal that the monolayer FeSe on an STO substrate is
heavily electron doped such that the Fermi surface consists of only electron pockets at Brillouin zone (BZ)
corners~\cite{Liu2012,He2013,Tan2013,Lee2014}. It is generally believed that this electron doping is caused
by oxygen vacancies in the STO surface induced by annealing of the substrate before the growth of
FeSe~\cite{Tan2013}. The large electron doping and the resulting Fermi surface with only electron pockets
directly challenge the Fermi-surface-nesting driven, purely electronic pairing
mechanism~\cite{Mazin2008a,Mazin2011}. One way to reconcile such a scenario is by taking account of the
holelike band located below the Fermi level at the $\Gamma$ point~\cite{Bang2014,Chen2015}, since the band
top is less than $100\ut{meV}$ away from the Fermi level, i.e., an ``incipient'' band~\cite{Chen2015} that
might fall in the low-energy cutoff of the bosons mediating pairing. The substrate, however, also influences
the phononic degrees of freedom and the electron-phonon ({\ep}) interaction, which can lead to phonon
contributions to the superconductivity. For example, the STO substrate has a stabilizing effect for the
sheering motion of the FeSe layer that serves to enhance the total coupling to the Fe and Se derived phonon
modes~\cite{Coh2015}. Another intriguing possibility is the presence of cross-interface coupling between the
FeSe layer and the substrate~\cite{Lee2014,Lee2015}. Evidence for the latter possibility has been inferred
from recent ARPES experiments (Ref.~\onlinecite{Lee2014} for FeSe/STO and later Ref.~\onlinecite{Peng2014a}
for FeSe/BTO), which observed replica bands $100\ut{meV}$ below the main electronic
bands~\cite{Lee2014,Peng2014a}. These replica bands were interpreted as shake-off states produced by the
coupling between the FeSe electrons and an oxygen optical phonon branch in the
substrate~\cite{Lee2014,Lee2015}.

The shape and intensity of the replica bands have been used to infer a strong coupling between the oxygen
optical phonons and the Fe $3d$ electrons that is strongly peaked for small momentum transfers (forward
scattering)~\cite{Lee2014}. This is a significant experimental result, as such coupling can produce
substantial enhancements in the total $T_c$ of FeSe/STO, even in unconventional channels where phonons are
not expected to play an essential role~\cite{Lee2014,Rademaker2016,Li2015c}. Moreover, this cross-interface
coupling provides a natural framework for understanding the $T_c$ enhancement in the FeSe/BTO
system~\cite{Peng2014a}. This momentum structure has been qualitatively confirmed by recent {\it ab initio}
calculations for the interface {\ep} coupling~\cite{Li2014}. However, in light of the sharpness of the
replica bands, the $q$-resolution in Ref.~\onlinecite{Li2014} is not in line with the sharpness of the
coupling in momentum space that is necessary to explain this experiment.

Motivated by this, we investigate the {\ep} coupling for films of FeSe on different oxide substrates using
{\it ab initio} methods. We first determine the phonon dispersion relations, the Eliashberg spectral function
$\alpha^2F(\omega)$, and the total coupling strength $\lambda$ for the following four systems: case (a), a
monolayer of FeSe on a {\STO} substrate (FeSe/STO); case (b), a monolayer of FeSe on an oxygen-vacant {\STO}
substrate with a $(1\times2)$ reconstruction (FeSe/STO1x2); case (c), a bilayer of FeSe on a {\STO} substrate
(2L-FeSe/STO); and case (d), a monolayer of FeSe on a {\BTO} substrate (FeSe/BTO). Next, we study the
momentum-dependent coupling strength $\lambda_{\nu\vq}$ for various phonon branches $\nu$ and specifically
focus on the topmost branch, an oxygen phonon branch whose energy coincides with the offset energy of the
replica bands measured by ARPES. Using a similar $q$ sampling, we find all four systems have a comparable
total coupling strength $\lambda=0.2\text{--}0.3$, consistent with the calculation result for the FeSe/STO in
Ref.~\onlinecite{Li2014}. Furthermore, for all systems and substrates explored, we investigate the momentum
dependence of the coupling in Sec.~\ref{sec:lambdaq} by computing the matrix elements $g(\vk,\vq)$ at a few
$q$ points at and very close to $\Gamma$, while, as expected from the experiments, we find a finite coupling
at $\vq=0$ and negligible couplings at $\vq\neq 0$ in our calculations. We find, however, that the integrated
coupling strength is insufficient to account for the high $T_c$ observed in the monolayer FeSe/STO system on
its own. This points to a cooperative role played by the cross-interface coupling. We discuss in the end a
few scenarios where the coupling to this branch may be enhanced and the necessity of a cooperative pairing
mechanism between the forward scattering {\ep} interaction and a purely electronic interaction based on the
current results.

\section{Model and Method}
\begin{figure}
  \centering
  \includegraphics[width=0.9\columnwidth]{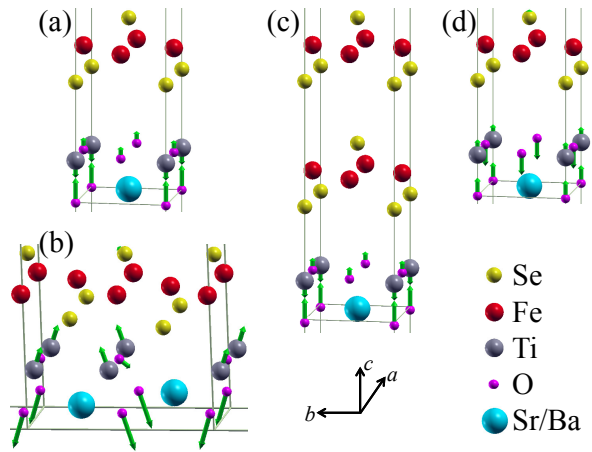}

\caption{(color online). Relaxed crystal structures for FeSe on TiO$_2$-terminated {\STO} [{\BTO} in (d)]. We
consider monolayers of FeSe in (a), (b), and (d) and a bilayer of FeSe in (c). The arrows indicate the
direction and amplitude of the atomic displacements associated with mode $\nu=27$ (a), $\nu=48$ (b), $\nu=39$
(c), and $\nu=24$ (d) (defined in text). The displacements are vectors $\mathbf{u}_{s,\nu}(\vq)$ with
components $u_{s,\nu}^{\alpha}(\vq) = \varepsilon_{s,\nu}^{\alpha}(\vq)/\sqrt{M_s}$, where
$\varepsilon_{s,\nu}^{\alpha}(\vq)$ are eigenvectors defined in the text. All structures are in a slab
geometry where the lattice is repeated in the $ab$-plane and separated by a vacuum layer before being
repeated in the $c$-direction. The vacuum layer is not shown here.}
 \label{fig:FeSe_structs}
\end{figure}

The four crystal structures considered in this work, each in a slab geometry, are shown in
Fig.~\ref{fig:FeSe_structs}. In all cases the substrates are one unit cell thick, and terminated at the
TiO$_2$ surface. We set $a=3.905\ut{\AA}$ as the in-plane lattice constant for both the substrate and FeSe
layer and place a vacuum layer around $12\ut{\AA}$ in height above the FeSe layers before the structure is
repeated in the $c$ direction. All structures are relaxed until a force smaller than $0.2\ut{meV/\AA}$ is
found on each atom. The relaxed structures for all cases are shown in Fig.~\ref{fig:FeSe_structs}. In
Fig.~\ref{fig:FeSe_structs}(b), the oxygen atoms in the Ti-O chain along the $b$ direction have been removed,
resulting in the more stable $(1\times2)$ reconstructed structure, similar to that inferred from several
experiments~\cite{Wang2012a,Li2014a} and a theoretical calculation~\cite{Bang2013a}. Thus, the lattice
constant in the $b$ direction is doubled $b=2a$ in case (b).

The electronic structure calculations are based on density functional theory (DFT) as implemented in the
Quantum \textsc{espresso} package~\cite{Giannozzi2009}. The exchange-correlation functional was taken in the
generalized gradient approximation (GGA) of Perdew-Burke-Ernzerhof (PBE) type~\cite{Perdew1996} within a
plane-wave pseudopotential representation. We choose an energy cutoff of $40\ut{Ry}$ ($50\ut{Ry}$) for plane
waves and $480\ut{Ry}$ ($700\ut{Ry}$) for charge densities (the higher energy cutoffs are for oxygen-vacant
FeSe/STO1x2). A $16\times16\times1$ ($16\times8\times1$ for FeSe/STO1x2) Monkhorst-Pack $k$ grid is used for
BZ summations in the DFT calculations. We have checked the computation results with the local-density
approximation (LDA) functional, $32\times32\times1$ $k$-grid sampling, or higher energy cutoffs, and found no
qualitative changes to our conclusions.

The dynamical properties of the lattice, including dynamical matrices, phonon dispersions, phonon density of
states (PDOS), and {\ep} coupling matrix elements and coupling strengths, are calculated with the
density-functional perturbation theory~\cite{Baroni2001} (DFPT) implemented in Quantum \textsc{espresso}. The
dynamical matrices are calculated on a $4\times4\times1$ $q$-grid ($4\times2\times1$ for FeSe/STO1x2) and
then Fourier transformed to force constants in real space. The $\nu$-th phonon mode of frequency
$\omega_{\vq\nu}$ at any wave vector $\vq$ is then calculated using the Fourier interpolation of the
dynamical matrices through the force constants, which is a standard technique~\cite{Gonze1997,Baroni2001}.

The {\ep} coupling function matrix elements are
\begin{align}
  g_\nu(i\vk,j\vk') &=
    \left(\frac{\hbar}{2\omega_{\nu \vq}}\right)^{1/2}
    \la \psi_{i\vk}| \Delta V_\text{SCF}(\nu \vq)|\psi_{j\vk'}\ra,
  \label{eq:gijkk}
\end{align}
where $\vq=\vk'-\vk$, $|\psi_{i\vk}\ra$ is the single-particle Bloch state characterized by wave vector $\vk$
and band index $i$ (here we explicitly write out the band index $i$ instead of suppressing it in $\vk$), and
\begin{align}
  \Delta V_\text{SCF}(\nu \vq) &= \sum_{\vR_{l}} \frac{e^{i\vq\cdot\vR_{l}}}{\sqrt{N}}  \sum_{s,\alpha}
    \frac{1}{\sqrt{M_s}}\frac{\partial V_\text{SCF}}{\partial \xi_{s}^{\alpha}}\varepsilon_{s,\nu}^{\alpha}(\vq),
  \label{eq:dVscf}
\end{align}
is the self-consistent first-order variation of the Kohn-Sham potential due to the small displacement
$\xi_{s}^{\alpha}$ of atom $s$ in the direction $\alpha$ of Cartesian coordinates. Here,
$\varepsilon_{s,\nu}^{\alpha}(\vq)$ is the eigenvector of the phonon mode with wave vector $\vq$, branch
index $\nu$, and frequency $\omega_{\nu\vq}$; $N$ is the number of unit cells in the crystal; $M_s$ is the
mass of atom $s\in\{1,\dots,S\}$ in the unit cell $\vR_{l}\in\{\vR_{1},\dots,\vR_{N}\}$; and
$\alpha\in\{1,2,3\}$ is the Cartesian coordinate index.

The dimensionless momentum-resolved coupling strength is defined as
\begin{align}
  \lambda_\nu(i\vk,j\vk+\vq) &=2N_\text{F}|g_\nu(i\vk,j\vk+\vq)|^2/\omega_{\nu \vq},
  \label{eq:modelambdakk}
\end{align}
where $N_\text{F}$ is the electronic density of states (DOS) per spin per unit cell at the Fermi level. The
dimensionless ``monochromatic'' coupling strength is defined as
\begin{align}
  \lambda_{\nu\vq} &= \frac{1}{N^2_\text{F}N}\sum_{\vk,ij}\lambda_\nu(i\vk,j\vk+\vq)
    \delta(\epsilon_{i\vk})\delta(\epsilon_{j\vk+\vq}),
  \label{eq:lambdaq}
\end{align}
and the total dimensionless {\ep} coupling constant (EPC) is defined as
\begin{align}
  \lambda &=  \frac{1}{N}\sum_{\vq,\nu} \lambda_{\nu\vq}.
  \label{eq:totlambda}
\end{align}
For the discussion below, denote $\lambda_{\vq} = \sum_{\nu}\lambda_{\nu\vq}$ and $\lambda_{\nu} =
\frac{1}{N}\sum_{\vq}\lambda_{\nu\vq}$. The Eliashberg spectral function is
\begin{align}
  \alpha^2 F(\omega) &= \frac{1}{2N}\sum_{\vq,\nu} \lambda_{\vq\nu}\omega_{\vq\nu}
    \delta(\omega-\omega_{\vq\nu}),
\end{align}
the frequency-dependent EPC is
\begin{align}
  \lambda(\omega) = 2\int_{0}^{\omega}\frac{\alpha^2F(\omega')}{\omega'}d\omega',
\end{align}
and the total EPC is $\lambda=\lambda(\infty)$. Last, the phonon density of states (PDOS) is $D(\omega) =
\frac{1}{N}\sum_{\vq\nu} \delta(\omega-\omega_{\vq\nu})$.

Before continuing, it should be noted that Eq.~(\ref{eq:lambdaq}), widely used in \textit{ab initio} studies,
is an approximate formula because the phonon energy transfer $\omega_{\nu\vq}$ has been dropped in one of the
delta functions in what should be an energy-conserved scattering process (the so called double-delta-function
approximation). If $\omega_{\nu\vq}$ is not small, the approximate result from Eq.~(\ref{eq:lambdaq}) will
deviate from the more accurate formula, especially for optical phonons with a finite coupling at
$\vq=0$~\cite{Calandra2005}. If only the total EPC is needed and the $\vq=0$ is assumed to have negligible
weight \textit{a priori}, one can apply Eq.~(\ref{eq:lambdaq}) but the convergence of the total EPC on the
$q$ grid needs to be checked. Because of this, a dense $q$ grid is necessary to accurately sum over $\vq$ in
Eq.~(\ref{eq:totlambda}).  For our systems it is impractical to directly calculate the coupling matrix
elements on every $q$ point in such a dense $q$ grid. Various interpolation techniques are available to
circumvent this difficulty, such as Fourier interpolation by maximally localized Wannier
functions~\cite{Giustino2007,Noffsinger2010,Calandra2010} or by using the auxiliary phonon
linewidths~\cite{Wierzbowska2005}, and an improved tetrahedron method~\cite{Kawamura2014}. We use the method
in Ref.~\onlinecite{Wierzbowska2005} (as it is already implemented in the Quantum \textsc{espresso} package)
to compute $\lambda_\vq$ on a dense $24\times24\times1$ $q$ grid that is needed for the summation in
Eq.~(\ref{eq:totlambda}). A $32\times32\times1$ $k$-grid and a broadening $\eta=0.005\ut{Ry}$ is used in
Eq.~(\ref{eq:lambdaq}). Note, however, that none of the interpolation techniques mentioned above can properly
treat the matrix elements with long spatial decay in real space, or with a sharp peak near $\vq=0$ in
momentum space~\cite{Sjakste2015,Verdi2015}. This is most likely to be the case where the $\vq=0$ weight is
not negligible. We will come back to this comment again below.

\section{Results and Analysis}

\subsection{Band Structure, Phonon Dispersion and {\ep} Coupling}
\begin{figure}
  \centering
  \includegraphics[width=1\columnwidth]{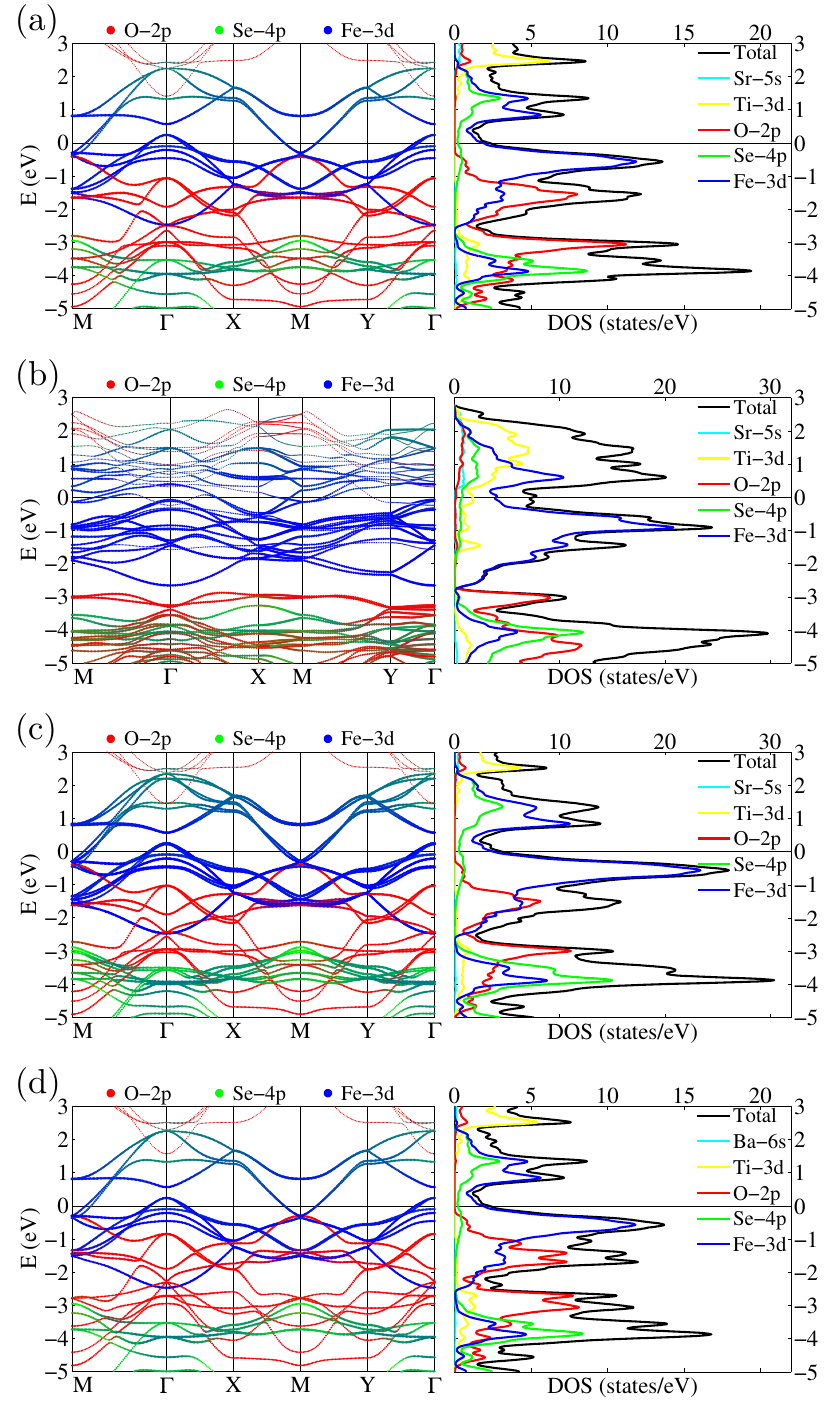}

\caption{(color online). (a) FeSe/STO, (b) oxygen-vacant FeSe/STO1x2, (c) 2L-FeSe/STO, (d) FeSe/BTO. Left:
band structure. The RGB values of the dots are determined by the orbital weight: red for O atoms, green for
Se atoms, and blue for Fe atoms. The size of the dots is proportional to the sum of the orbital weight
considered. Right: total and projected DOS (summed for two spin components).}
 \label{fig:ebnd}
\end{figure}
\begin{figure}
  \centering
  \includegraphics[width=1\columnwidth]{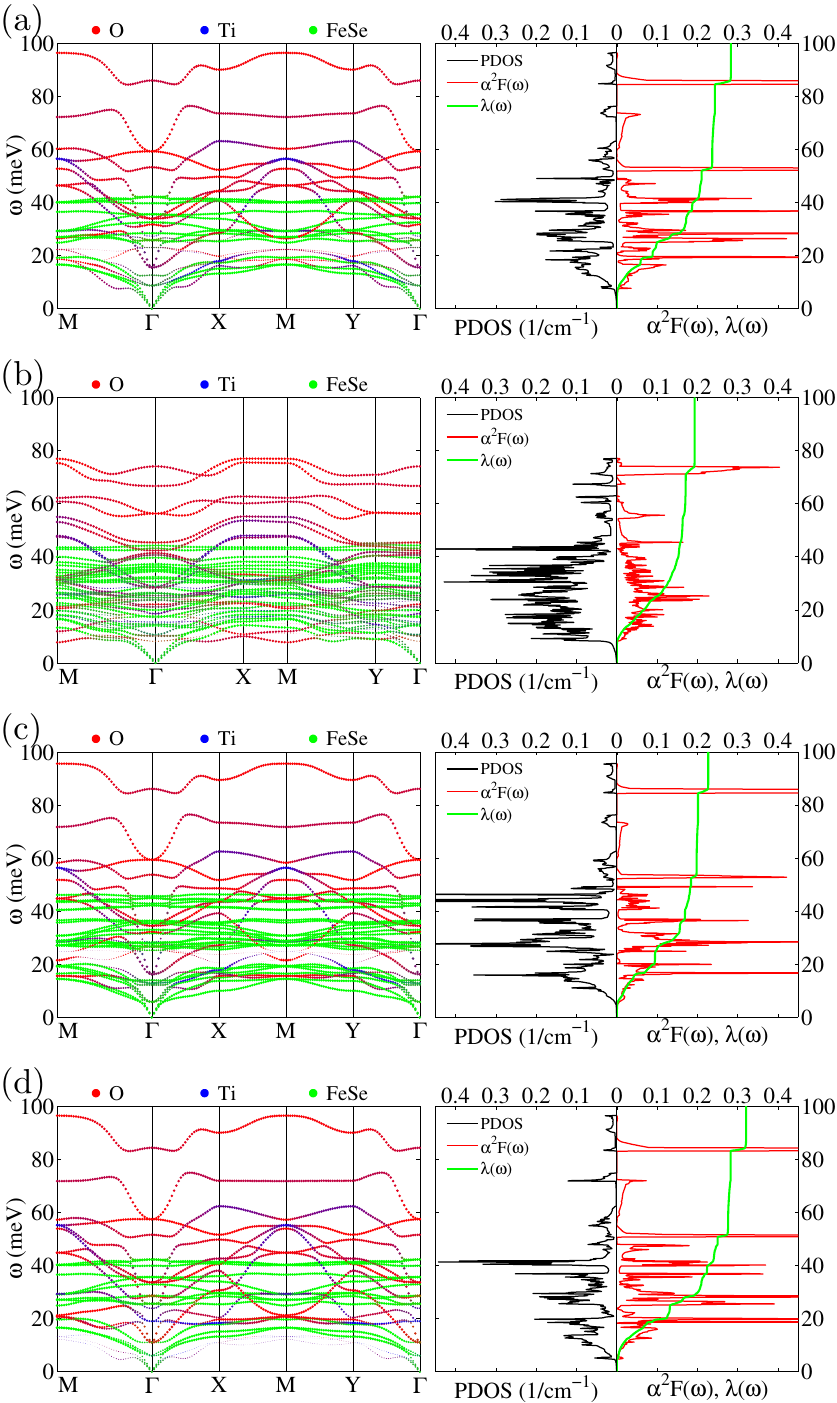}

\caption{(color online). (a) FeSe/STO, (b) oxygen-vacant FeSe/STO1x2, (c) 2L-FeSe/STO, (d) FeSe/BTO. Left:
phonon dispersion. The RGB values of the dots are determined by the eigenvector
$|\varepsilon_{s,\nu}^{\alpha}(\vq)|^2$ of the mode $\omega_{\nu\vq}$: red for O atoms, blue for Ti atoms,
and green for both Fe and Se atoms. The size of the dots is proportional to the sum of the eigenvector
components considered. Right: phonon density of states (black line), Eliashberg spectral function
$\alpha^2F(\omega)$ [red (gray) line] and {\ep} coupling $\lambda(\omega)$ [green (light gray) line].}
 \label{fig:phbnd}
\end{figure}

Figure \ref{fig:ebnd} shows the calculated electronic band structure and DOS for the corresponding four cases
defined before. The results in Fig.~\ref{fig:ebnd}(a) and Fig.~\ref{fig:ebnd}(b) agree well with prior
calculations~\cite{Li2014,Bang2013a}. The bands near the Fermi level are mainly Fe $d$ bands. Figure
\ref{fig:ebnd}(a) shows a peak in the O $p$ density of states around $-2\ut{eV}$ which however is absent in
the presence of oxygen vacancies as shown in Fig.~\ref{fig:ebnd}(b). The similarity of band structure in
Fig.~\ref{fig:ebnd}(a) and Fig.~\ref{fig:ebnd}(d) indicates that there is only a small difference between STO
and BTO substrate in electronic structure; Fig.~\ref{fig:ebnd}(c) shows that the second FeSe layer simply
doubles the Fe $d$ states near the Fermi level. In Fig.~\ref{fig:ebnd}(b), however, the O vacancy strongly
changes the band structure by electron-doping the system and removing the Fe $d$ hole pockets at $\Gamma$. In
addition, there is an increment of Ti $d$ states near the Fermi level and a quite large Ti $d$ electron
pocket around $\Gamma$.

Figure \ref{fig:phbnd} shows the phonon dispersion relations, the PDOS, and the EPC
$\lambda=\lambda(\infty)$. One important result is that the topmost phonon branch (denoted as mode $\nu$
below)---a set of oxygen polar modes---has an energy around $80\text{--}100\ut{meV}$ in each case. The
dispersion of this branch softens to an energy below $80\ut{meV}$ in case (b) for the system with O-vacant
substrate, which can be explained by the charge transfer between the STO substrate and FeSe
monolayer~\cite{Xie2015}. The phonon energy of this oxygen phonon is consistent with the inferred $\sim
80\ut{meV}$ phonon that causes the replica bands seen in ARPES experiments~\cite{Lee2014}. The displacement
pattern of the $\vq = 0$ mode is shown in Fig.~\ref{fig:FeSe_structs} for each case. These vibrations can
induce excess $z$-directional dipole moments situated at a plane near the surface, as suggested in
Refs.~\onlinecite{Lee2014,Lee2015}, and result in an {\ep} coupling between the substrate phonons and the
FeSe electrons. Indeed, we find that this mode alone contributes a sizable amount to the total {\ep} coupling
strength and has a relatively flat dispersion, resulting in a sharp peak in the Eliashberg spectral function
$\alpha^2F(\omega)$ (right panels in Fig.~\ref{fig:phbnd}). Our result shows that this peak is not unique to
the FeSe/STO system~\cite{Li2014}, but also exists in BTO and/or O-vacant STO substrates. Further
investigation, presented in the next section, shows that most of the coupling to this oxygen mode is due to
the intraband matrix elements of zero momentum transfer.

We stress that there are two factors that contribute to an overestimation of $\lambda$. First, the standard
interpolation techniques do not work properly when the coupling is strongly peaked at $\vq=0$, as they tend
to overestimate the width of the peak in momentum space. Second, this mode has a finite contribution at
$\vq=0$, which will be strongly boosted by the double-delta-function approximation. The net result is that
the contribution of this coupling to total EPC $\lambda(\infty)$ in Fig.~\ref{fig:phbnd} is overestimated and
should be considered as an upper limit of this branch's contribution to the total EPC, as determined by first
principles. For the same reasons, the calculated coupling strengths and widths in $q$ space in
Ref.~\onlinecite{Li2014} are overestimated.

Before we introduce a proper way to characterize the $q$ dependence of the {\ep} interaction without
resorting to the double-delta-function approximation, we compare some other aspects of the four cases. First,
from the color-coded dispersion relations, we see that the phonon modes of the FeSe layer are all below
$50\ut{meV}$. We again find the two systems---(a) FeSe/STO and (d) FeSe/BTO---have a similar phonon spectrum,
PDOS, and EPC. [One difference is that the Ba atom has a large weight in the eigenvector of the lowest modes
in panel (d) as the weights plotted for O, Ti, and FeSe are small.] Comparing Fig.~\ref{fig:phbnd}(a) and
Fig.~\ref{fig:phbnd}(c), we find that the second layer of FeSe does not alter the existing phonon modes too
much and seems to only add more phonon modes below $50\ut{meV}$. Nevertheless, the corresponding {\ep}
coupling is smaller. Finally, the total EPC for case (b) is also smaller than the other cases. We summarize
the quantitative results in Table~\ref{tab:lambda}. In the table, we define $\omega_{\ln} =
\exp\left(\frac{2}{\lambda} \int d\omega \omega^{-1}\alpha^2F(\omega)\ln\omega\right)$ and
$\int_0^{\omega_D}D(\omega) d\omega = 3S/2$, where $D(\omega)$ is the phonon DOS, $3S$ is total number of
phonon modes, and $\omega_D$ is the Debye energy defined here.

\begin{table}

\caption{Numerical values of the calculation result for all cases: (a) FeSe/STO, (b) oxygen-vacant
FeSe/STO1x2, (c) 2L-FeSe/STO, and (d) FeSe/BTO. The DOS $N_\text{F}$ is per spin per unit cell. The
particular oxygen mode is $\nu=27$ [(a), (d)], $\nu=48$ (b), and $\nu=39$ (c).} \label{tab:lambda}
\renewcommand{\arraystretch}{1.4}
\begin{tabular*}{\columnwidth}{@{\extracolsep{\fill}}cccccc}
\hline \hline
Case &$\lambda$ &$\lambda_{\nu}$ &$N_\text{F}\ut{(states/eV)}$
     &$\omega_\text{ln}\ut{(meV)}$  &$\omega_D\ut{(meV)}$ \\\hline
(a)  &0.283     &0.040      &1.00       &29.8       &31.0       \\
(b)  &0.193     &0.022      &3.67       &25.9       &27.8       \\
(c)  &0.227     &0.025      &2.02       &25.9       &31.1       \\
(d)  &0.321     &0.038      &1.00       &27.7       &30.0       \\
\hline \hline
\end{tabular*}
\end{table}

\subsection{Momentum Dependence of {\ep} Coupling} \label{sec:lambdaq}
When we consider the momentum $\vq$ dependence of the {\ep} coupling as defined in Eq.~(\ref{eq:lambdaq}), we
find a few disadvantages that are related to the double-delta-function approximation that we mentioned
before. First, the two delta functions in Eq.~(\ref{eq:lambdaq}) require a large $k$-point sampling to
achieve an accurate result, so the $k$-summed coupling strength is sensitive to $k$ grid and Fermi surface
broadening. Second, the nesting property of the Fermi surface, i.e., the phase space for scattering on the
Fermi surface, will strongly affect the value of the two delta functions and make it difficult to infer or
compare the magnitude of the coupling matrix elements $|g_\nu(i\vk,j\vk+\vq)|^2$ or
$\lambda_\nu(i\vk,j\vk+\vq)$ from the calculated $\lambda_{\nu\vq}$.

In order to circumvent these difficulties, we define a Fermi surface average $\la \lambda_{\nu\vq} \ra$
by separating the nesting property and the matrix elements in $\lambda_{\nu\vq}$. We begin with the phonon
linewidth~\cite{Grimvall1981,Giustino2007}
\begin{align}
  &\gamma_{\nu\vq} = -\Im \Pi_{\nu\vq} = \frac{2\pi\omega_{\nu\vq}}{N}\sum_{\vk,ij}|g_\nu(i\vk,j\vk+\vq)|^2 \notag\\
  &\times \frac{f(\epsilon_{i\vk})-f(\epsilon_{i\vk}+\omega_{\nu\vq})}{\omega_{\nu\vq}}
   \delta(\epsilon_{j\vk+\vq}-\epsilon_{i\vk}-\omega_{\nu\vq}),
  \label{eq:phlinw}
\end{align}
where $\Pi_{\nu\vq}$ is the phonon self-energy and $f(x)=1/[\exp(x/T)+1]$ is the Fermi distribution function.
We have replaced $\epsilon_{j\vk+\vq}$ in the second Fermi distribution with
$\epsilon_{i\vk}+\omega_{\nu\vq}$. The ``monochromatic'' coupling strength is then given by
\begin{align}
  \tilde{\lambda}_{\nu\vq} &= \frac{\gamma_{\nu\vq}}{\pi N_\text{F} \omega^2_{\nu\vq}}.
  \label{eq:lmdlinw}
\end{align}
Only when $\omega_{\nu\vq}$ is much smaller than the temperature broadening, do we have
$\tilde{\lambda}_{\nu\vq} \approx \lambda_{\nu\vq}$. Next, we define the nesting
function~\cite{Krakauer1993,Kasinathan2006}
\begin{align}
  \tilde{\xi}(\vq,\omega_{\nu\vq}) = &
  \frac{1}{N}\sum_{\vk,ij} \frac{f(\epsilon_{i\vk})-f(\epsilon_{i\vk}+\omega_{\nu\vq})}{\omega_{\nu\vq}} \notag \\
  &\times \delta(\epsilon_{j\vk+\vq}-\epsilon_{i\vk}-\omega_{\nu\vq}),
  \label{eq:nestfunlinw}
\end{align}
and the approximate form
\begin{align}
  \xi(\vq) = \frac{1}{N}\sum_{\vk,ij}\delta(\epsilon_{i\vk}) \delta(\epsilon_{j\vk+\vq}).
  \label{eq:nestfun}
\end{align}
Some of the properties of $\xi(\vq)$ are discussed in Ref~\onlinecite{Kasinathan2006}. It is easy to see that
$N^{-1}\sum_{\vq}\xi(\vq)=N^2_\text{F}$. Finally, we define the new $k$-averaged coupling constants as
\begin{align}
  \la\lambda_{\nu\vq}\ra &= \frac{N^{2}_\text{F}\lambda_{\nu\vq}}{\xi(\vq)}, \\
  \la \tilde{\lambda}_{\nu\vq}\ra &= \frac{N^{2}_\text{F}\tilde{\lambda}_{\nu\vq}}{\tilde{\xi}(\vq,\omega_{\nu\vq})}.
\end{align}
These coupling constants characterize the $\vq$ dependence of the {\ep} matrix element
$|g_\nu(i\vk,j\vk+\vq)|^2$, independently of the Fermi surface shape and the size of phase space for
scattering processes determined by the Fermi surface shape. The tilde ($\tilde{\ }$) indicates including the
phonon frequency in one of the delta functions, while the nontilde notation means that the
double-delta-function approximation is applied.

\begin{figure}
  \centering
  \includegraphics[width=1\columnwidth]{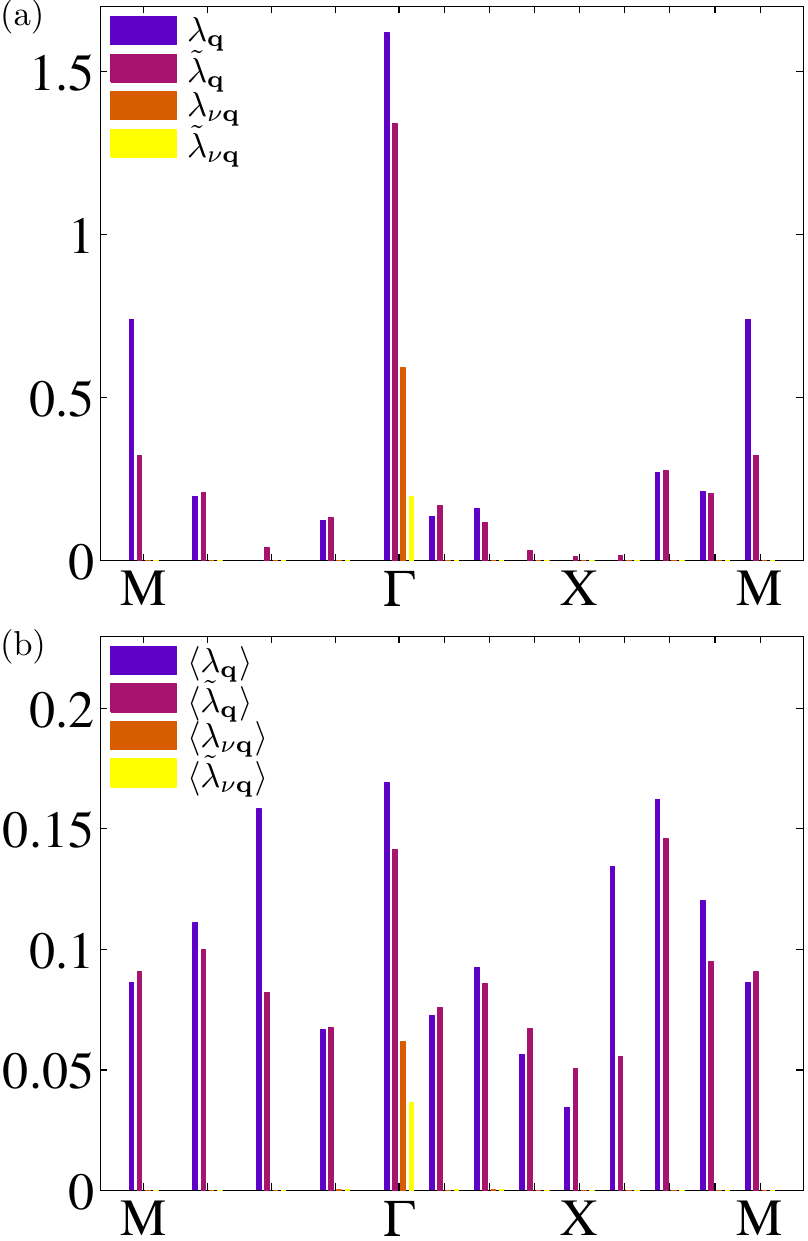}

\caption{(color online). The $q$ dependence of the {\ep} coupling strength $\lambda_\vq$ for the FeSe/STO
system. $\lambda_{\nu\vq}$ for topmost mode $\nu=27$ is also shown. (a) Coupling strength calculated with
approximated ($\lambda$) and exact formula ($\tilde\lambda$). (b) Averaged coupling strength, i.e., coupling
strength in (a) normalized by the corresponding nesting function.}
 \label{fig:qlmd}
\end{figure}

In Fig.~\ref{fig:qlmd}(a), we plot the calculated $\lambda_{\nu\vq}$ and the mode-summed
$\lambda_\vq=\sum_\nu \lambda_{\nu\vq}$, using a denser $8\times 8 \times 1$ $q$ grid to illustrate the
momentum dependence of the interaction. In Fig.~\ref{fig:qlmd}(b), we plot $\la\lambda_{\nu\vq}\ra$ and
$\la\lambda_\vq\ra$, which were computed using the {\ep} coupling matrix elements $g_\nu(i\vk,j\vk+\vq)$
directly calculated by DFPT at each momentum $\vq$ with a $16\times 16\times 1$ $k$ grid and then
interpolated to a $64\times 64\times 1$ $k$ grid for the $k$ sum. Only the bands $i,j$ crossing the Fermi
level are included in the sum. The delta function is approximated by a Gaussian $\delta(x) =
\frac{1}{\sqrt{\pi}\eta} e^{-x^2/\eta^2}$. The temperature broadening in the Fermi distribution function and
Gaussian broadening in the delta function are both set to $0.005\ut{Ry}$.

In Fig.~\ref{fig:qlmd}(a), we see that the mode-summed couplings have a very strong $q$ dependence, whether
we use exact Eq.~(\ref{eq:lmdlinw}) or approximate Eq.~(\ref{eq:lambdaq}); however, in
Fig.~\ref{fig:qlmd}(b), the mode-summed couplings all reach a comparable level across the high-symmetry path
when the size of phase space for scattering processes is separated by the normalization of the nesting
function. This indicates that the total $e$-$ph$ interaction, averaged over all modes, is fairly momentum
independent. In contrast, $\la\lambda_{\nu\vq}\ra$ (and $\la\tilde{\lambda}_{\nu\vq}\ra$) of the oxygen mode
$\nu=27$ (for the FeSe/STO system) is still peaked at $\vq=0$. (Mode $\nu=24$, which corresponds to the
optical oxygen branch at $\sim 60$~meV, shows similar behavior but is not plotted.) Since
$\la\lambda_{\nu\vq}\ra$ truly reflects the magnitude of the matrix elements near the Fermi surface, the
matrix elements (computed within Quantum \textsc{espresso}) must also peak at $\vq=0$ and decay very fast
away from it.

\begin{figure}
  \centering
  \includegraphics[width=0.97\columnwidth]{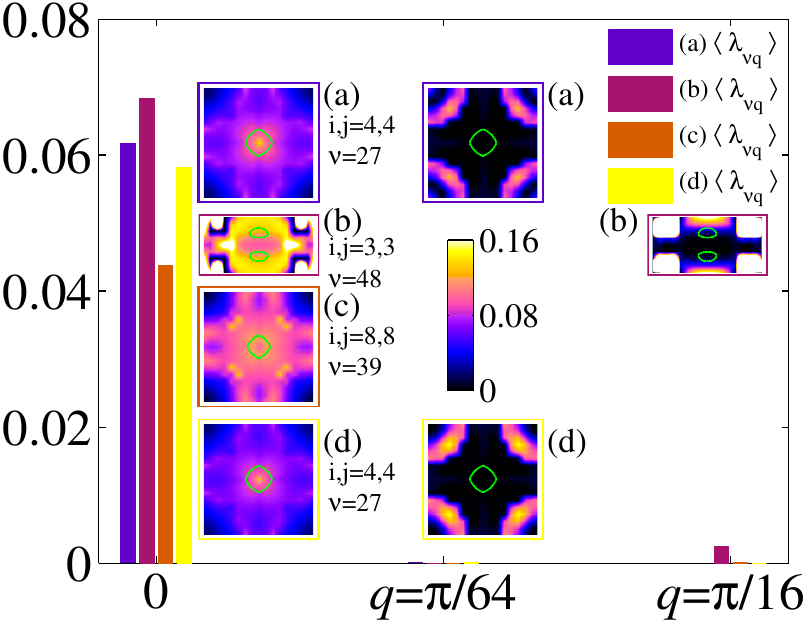}

\caption{(color online). The coupling strength $\la\lambda_{\nu\vq}\ra$ obtained by averaging the matrix
elements (bar graph) and intraband coupling matrix elements $\lambda_{\nu}(i\vk,j\vk+\vq)$ in the first
Brillouin zone (the rectangular inset panels) for a few $q$ points near $\vq=0$ for four cases: (a) FeSe/STO,
(b) oxygen-vacant FeSe/STO1x2, (c) 2L-FeSe/STO, and (d) FeSe/BTO. Here, $\vq=(0,q,0)\frac{1}{a}$ and mode
$\nu$ is the topmost mode in the dispersion for each case. Only one pair of $(i,j)$ bands is shown for each
case. The corners of each inset are $\Gamma$ points and the center $M$ point. The magnitude of the matrix
elements is indicated by the colorbar with darker color for lower value and brighter color for higher value
(any out-of-limit value is indicated by black or white). The green solid line in the inset panels is the
electron pocket for the corresponding band. The matrix elements plotted in the inset panels are multiplied by
$400$ at $q_y=\pi/64$ and by $10$ at $q_y=\pi/16$.}
 \label{fig:qpathlmd}
\end{figure}

We can verify this in Fig.~\ref{fig:qpathlmd}, where we plot $\la\lambda_{\nu\vq}\ra$ and the relevant matrix
elements at a few selected $q$-points very close to $\vq=0$. [We calculate $q=(0,0,0)$ and $(0,\pi/64,0)$ for
case (a) and (d); $q=(0,0,0)$ and $(0,\pi/16,0)$ for case (b); but only $q=(0,0,0)$ for case (c) because of
the difficulty of convergence in case (c) for $q$ very close to the $\Gamma$ point.] Here, results are again
shown for the topmost phonon mode of each case that we considered. Note, $\la\lambda_{\nu\vq}\ra$ is the
average of corresponding matrix elements (insets in Fig.~\ref{fig:qpathlmd}) summed over different bands; by
definition only the matrix elements near the Fermi surface contribute to the average, and the size of the
phase space for the scattering processes does not affect $\la\lambda_{\nu\vq}\ra$ because it is normalized
with respect to the nesting function. Although it is consistent with values on the coarse $q$ grid shown in
Fig.~\ref{fig:qlmd}, the matrix elements decay surprisingly fast away from $\vq=0$. The estimated peak width
is smaller than $\pi/64$, as shown by the bar graph of $\la\lambda_{\nu\vq}\ra$.
$\la\tilde{\lambda}_{\nu\vq}\ra$ is similar. This result shows that all four cases that we have considered
here have a (surprisingly) sharp $q$-dependent $e$-$ph$ coupling with the topmost oxygen optical phonons, and
that this interaction is strongly peaked at small momentum transfers. This result also demonstrates that the
width of this interaction in momentum space is much narrower than the one inferred in
Ref.~\onlinecite{Li2014} with the use of the double-delta-function approximation and problematic
interpolation methods.

Our calculated $q$ dependence of the coupling matrix elements is very sharp. We have cross-checked the
existence of the striking difference between coupling matrix elements at $\vq=0$ and $\vq\neq 0$ and found
that it is a consistent result, appearing when we use LDA, PBE, and PBEsol types of exchange-correlation
potentials in both norm-conserving and ultra-soft pseudopotential up to $\vq = (q_x,0,0)$ with $q_x$ as low
as $\frac{1}{200}\frac{2\pi}{a}$.

\begin{figure}
  \centering
  \includegraphics[width=0.97\columnwidth]{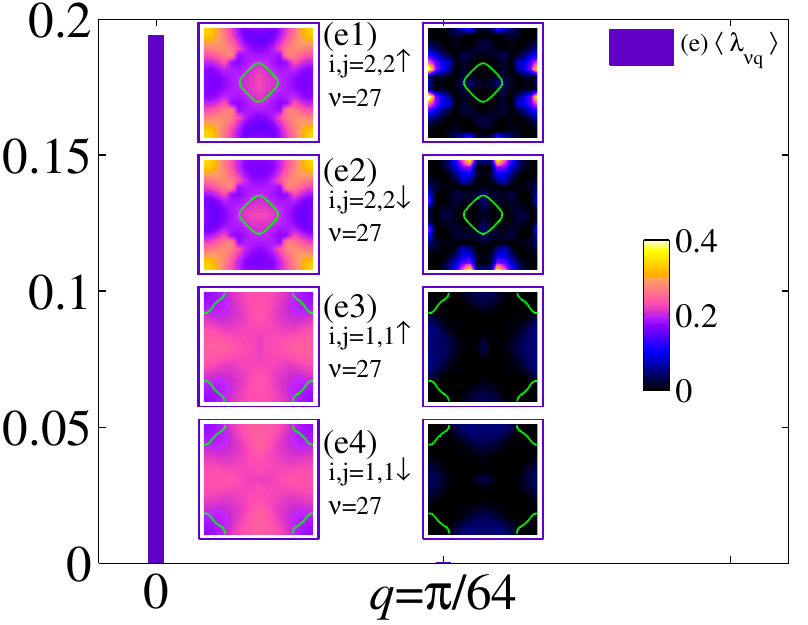}

\caption{(color online). The coupling strength $\la\lambda_{\nu\vq}\ra$ obtained by averaging the matrix
elements (bar graph) and intraband coupling matrix elements $\lambda_{\nu}(i\vk,j\vk+\vq)$ in the first
Brillouin zone (the inset panels) for two $q$-points near $\vq=0$ for cAFM FeSe/STO. Here,
$\vq=(0,q,0)\frac{1}{a}$ and mode $\nu$ is the topmost mode in the dispersion. (e1) and (e2) are for the
electron pocket at $M$ for spin-up and spin-down component, respectively. (e3) and (e4) are for the hole
pocket at $\Gamma$ for spin-up and spin-down component, respectively. The corners of each inset panel are
$\Gamma$ points and the center $M$ point. The green solid line in the inset panels is the Fermi pocket for
the corresponding band. The matrix elements plotted in the inset panels are multiplied by $2000$ at
$q_y=\pi/64$.}
 \label{fig:magqpathlmd}
\end{figure}

\section{Discussion}
$\la\lambda_{\vq}\ra$ at $\vq=0$ shown in Fig.~\ref{fig:qlmd}(b) indicates that it has a sizable contribution
to the total EPC $\lambda$. The approximate nesting function is also overestimated at $\vq=0$~\cite{Note1};
this is why $\lambda_{\vq}$ is so large in Fig.~\ref{fig:qlmd}(a). Due to the concurrence of these two
effects, the calculated total $\lambda = N^{-1}\sum_\vq \lambda_{\vq} = N^{-1}\sum_\vq \la\lambda_{\vq}\ra
N_\text{F}^{-2} \xi(\vq)$ could also be overestimated, depending on factors such as the density of the
$q$-grid and size of the broadening. Since in all the interpolation techniques mentioned before, the coupling
at $\vq=0$ is extrapolated to a finite region, it is important to resolve this region in the initial matrix
elements by direct calculation as we have shown in Fig.~\ref{fig:qpathlmd}. We have indeed found that the
calculated $\lambda$ for an $8\times 8 \times 1$ $q$-grid is smaller than that listed in
Table~\ref{tab:lambda} for a $4\times 4 \times 1$ $q$-grid. In either case, the calculated total {\ep}
coupling strength $\lambda=0.2\text{--}0.3$ for all four systems that we have considered cannot account for
the high $T_c\sim 70\ut{K}$ observed experimentally through the conventional phonon-mediated pairing
mechanism.

On the other hand, ARPES experiments~\cite{Lee2014,Peng2014a} found replica bands in the electronic structure
in these systems, suggesting a strong {\ep} coupling to phonons with mode energy $\sim
100\ut{meV}$.\cite{Lee2014,Lee2015} We have also found the suggested oxygen mode in our calculations for FeSe
on STO or BTO substrates. More importantly, as shown in Fig.~\ref{fig:qlmd} and Fig.~\ref{fig:qpathlmd}, all
of these systems have a sharp peak and a nonzero coupling strength near $\vq=0$; i.e., they favor the forward
scattering process. (Our calculated coupling, however, is much sharper in $q$ space than that estimated from
the experiments in Ref.~\onlinecite{Lee2014}.) The study in Ref.~\onlinecite{Rademaker2016} suggests that the
coupling to the forward scattering process results in a $T_c$ that depends linearly on the coupling constant,
where an estimated coupling strength $\sim 0.15\text{--}0.2$ for the single mode alone can account for the
total $T_c\sim 70\ut{K}$ if the Coulomb pseudopotential $\mu^*$ is neglected. Therefore, the $T_c$
enhancement due to this oxygen mode can be much larger than that expected from the conventional phonon
mediated pairing mechanism. However, the coupling strength we obtain here for coupling to this single oxygen
branch ($\lambda=0.02\text{--}0.04$) is still too small to account for the full $T_c$, even when we consider
the more effective pairing produced by the strong forward scattering nature of the interaction.

At this time there are a number of factors that could provide a satisfactory explanation for this
discrepancy. The first is that the {\ep} coupling can be enhanced when the correlations~\cite{Kulic1994} or
magnetic structure~\cite{Li2014} are considered. In Fig.~\ref{fig:magqpathlmd}, we plot $\la \lambda_{\nu\vq}
\ra$ (summed for two spins) and the spin-dependent matrix elements from the calculation for FeSe/STO with a
checkerboard antiferromagnetic (cAFM) spin configuration [denoted as the case (e)]. The electronic structure
(not shown) and the enhanced total {\ep} coupling constant $\lambda$ are consistent with the previous
calculation for the FeSe/STO case~\cite{Li2014}; however, $\la \lambda_{\nu\vq} \ra$ for topmost oxygen
branch decays at least as fast as in the other cases we have shown and therefore the integrated coupling for
this branch is still quite small, albeit with an increased coupling strength at $\vq=0$. Furthermore, there
are indications that strong electronic correlations can renormalize the {\ep} coupling preferentially at
small momentum transfers~\cite{Kulic1994,Huang2003}. In order to address this possibility extensions beyond
DFT are likely required~\cite{Coh2015,Li2015c}. Another possibility is that vertex corrections to the {\ep}
interaction, which were neglected in Ref.~\onlinecite{Rademaker2016}, may need to be included since they can
enhance $T_c$ in the perturbative regime when the {\ep} interaction is peaked at small momentum
transfers~\cite{Pietronero1995,Grimaldi1995,Kostur1994}. Our DFT results hint that FeSe/STO is in this regime
providing that a finite integrated coupling strength from the forward-focused coupling of the oxygen mode can
be obtained with an improved method. Another reason why the {\ep} interactions might be underestimated in our
simulations is that we only considered one STO layer instead of a semi-infinite number of STO layers.
Although their coupling to the Fe $d$ bands is expected to decay as one goes deeper into the STO substrate,
the sum of their contributions can still be significant especially for small momenta.

Another possible explanation is that the ferroelectric substrate and the two-dimensionality of our system
needs a more careful treatment that is beyond the current standard DFPT
routines~\cite{Sjakste2015,Sohier2015,Verdi2015}. For example, by proposing a charge depletion region across
multiple unit cells in the STO substrate near the interface, Zhou \textit{et al.}~\cite{Zhou2016} obtained
from calculations a much larger total coupling strength $\lambda\sim 0.4$ to the topmost oxygen phonon
branch, which is peaked at small $\vq$. Furthermore, the exact structure of the terminating layer of the
substrate has yet to be determined. One recent experiment~\cite{Li2016a} found the top two layers of the STO
substrate (prepared by Se etching) are two adjacent TiO$_2$ layers. If oxygen vibrations in both layers
contribute to the coupling to the $d$ electron in FeSe layer, a stronger coupling strength is expected.

Finally, the unconventional channel of electronic pairing mechanism can play an equal, if not larger, role in
the high $T_c$ in monolayer FeSe/STO or FeSe/BTO systems. There is growing experimental evidence for this
scenario. For example, the observation of superconductivity with $T_c\sim 40\ut{K}$ by field-effect
~\cite{Shiogai2016,Hanzawa2016,Lei2016} and potassium doping/surface coating~\cite{Miyata2015,Song2015} on
FeSe thin films supports this conclusion by indicating the action of an unconventional pairing mechanism.
However, without the STO or BTO substrates, the $T_c$ does not reach the value $\sim 70\ut{K}$, as shown by
the experiments on potassium surface coating on bulk FeSe crystal~\cite{Seo2015,Ye2015}. The presence of an
electronic pairing mechanism can also explain the fact that the bilayer FeSe/STO shows similar phonon
spectrum and {\ep} coupling strength, but does not superconduct in reality. Since the forward scattering
pairing is mainly intraband in nature, it can work in conjunction with the unconventional pairing mechanism
in most instances and explain the high $T_c$ observed in the monolayer FeSe systems with the cooperative
pairing mechanism.

\section{Conclusions}
We have calculated the phonon spectrum and {\ep} coupling strength for a monolayer and bilayer of FeSe on
pristine STO or BTO substrates or on an O-vacant STO substrate. We have found that an interfacial
$80\text{--}100\ut{meV}$ ferroelectric oxygen phonon branch couples to Fe $d$ electrons in all model
structures. The eigen displacements of this mode lead to a dipole potential scattering electrons with small
momentum transfer. The energy of this mode coincides with the offset of the replica bands measured in ARPES
and the coupling matrix elements have a sharp peak in $q$ space, preferring the forward scattering process.
The calculated coupling strength is insufficient to explain the high $T_c$ observed by ARPES experiments
through the phonon-mediated pairing mechanism for either the momentum-independent coupling or the forward
scattering coupling. Our results suggest that the inferred coupling enhances $T_c$ through a cooperative
mechanism with an unconventional pairing channel. Other types of structures with different terminating layers
of the substrate or more advanced treatment of the polar property of the ferroelectric substrate can possibly
lead to a moderate but sufficient coupling strength. Exploring these possibilities is left for future work.

\begin{acknowledgments}
We thank A.~Kemper for useful discussions. S.~J. and Y.~W. are partially funded by the University of
Tennessee's Science Alliance Joint Directed Research and Development (JDRD) program, a collaboration with Oak
Ridge National Laboratory. S.~J. acknowledges additional support from the University of Tennessee's Office of
Research and Engagement's Organized Research Unit program. A.~L. was supported by Grant
No.~DE-FG02-05ER46236. A portion of this research was conducted at the Center for Nanophase Materials
Sciences, which is a Department of Energy (DOE) Office of Science User Facility. This research used
computational resources supported by the University of Tennessee and Oak Ridge National Laboratory's Joint
Institute for Computational Sciences and resources of the National Energy Research Scientific Computing
Center (NERSC), a DOE Office of Science User Facility.
\end{acknowledgments}

\end{document}